\documentclass{nature}

\title{Nematic Fluctuations in Iron-Oxychalcogenide Mott Insulators}

\author{B. Freelon,$^{1,2}$ R. Sarkar,$^{3}$ S. Kamusella,$^{3}$ F. Br\"uckner,$^{3}$ V. Grinenko,$^{3,4}$ Swagata Acharya,$^{5}$ Mukul Laad,$^{6,7}$ Luis Craco,$^{8}$ Zahra Yamani,$^{9}$ Roxana Flacau,$^{10}$  Ian Swainson,$^{11}$ Benjamin Frandsen,$^{12}$ Robert Birgeneau,$^{12}$ Yuhao Liu,$^{13}$ Bhupendra Karki,$^{2}$ Alaa Alfailakawi,$^{2}$ Joerg C. Neuefeind,$^{14}$ Michelle Everett,$^{14}$ Hangdong Wang,$^{15}$ Binjie Xu,$^{15}$ Minghu Fang,$^{15,16}$ and H.-H. Klauss $^{3}$}
\usepackage[ansinew]{inputenc}
\usepackage{graphicx}
\usepackage{dcolumn}
\usepackage{bm}
\usepackage{xcolor}
\usepackage{stmaryrd}
\usepackage{amsmath, amsthm, amssymb}

\begin{document}
\newcommand{\cSe}{La$_{2}$O$_{2}$Fe$_{2}$OSe$_{2}$}
\newcommand{\cS}{La$_{2}$O$_{2}$Fe$_{2}$OS$_{2}$}
\newcommand{\cSSe}{La$_{2}$O$_{2}$Fe$_{2}$O(S, Se)$_{2}$}
\newcommand{\cSelayer}{Fe$_{2}$O\it{M}$_{2}$}
\newcommand{\muSR}{$\upmu$SR\xspace}
\newcommand{\mus}{$\upmu$s\textsuperscript{-1}\xspace}
\maketitle
\begin{affiliations}
 \item Department of Physics and Texas Center for Superconductivity, University of Houston, Texas 77004, USA
 \item Department of Physics, University of Louisville, Louisville, KY, 40208, USA,
 \item Institute of Solid State and Materials Physics, Technical University Dresden, 01062 Dresden, Germany
 \item IFW Dresden, 01171 Dresden, Germany
 \item Department of Physics, King's College London, London WC2R2LS United Kingdom
 \item Institute of Mathematical Sciences, Taramani, Chennai 600113, India
  \item Homi Bhabha National Institute Training School Complex, Anushakti Nagar, Mumbai 400085, India
 \item Instituto de F\'{\i}sica, Universidade Federal de Mato Grosso, 78060-900, Cuiab\'a, MT, Brazil
 \item Canadian Nuclear Laboratories, Chalk River Laboratories, Chalk River, Ontario K0J 1J0, Canada
 \item Physics Department, Santa Clara University, 301 Daly Science Center, Santa Clara, CA 95053
 \item Physics Section, International Atomic Energy Agency, Vienna International Centre,1400 Vienna, Austria
 \item Department of Physics, University of California, Berkeley, CA 94720
 \item Department of Materials Science and Engineering, University of Illinois Urbana Champaign, Urbana, Illinois, 61801, USA
 \item Neutron Scattering Division, Oak Ridge National Laboratory, Oak Ridge, TN, United States
 \item Department of Physics, Zhejiang University, Hangzhou 310027, P. R. China
\item  Collaborative Innovation Center of Advanced Microstructures, Nanjing University, Nanjing 210093, China
\end{affiliations}
  
\begin{abstract}
Nematic fluctuations occur in a wide range of physical systems from liquid crystals to biological molecules to solids such as exotic magnets, cuprates and iron-based high-\begin{math}T_c\end{math} superconductors. Nematic fluctuations are thought to be closely linked to the formation of Cooper-pairs in iron-based superconductors. It is unclear whether the anisotropy inherent in this nematicity arises from electronic spin or orbital degrees of freedom. We have studied the iron-based Mott insulators La$_{2}$O$_{2}$Fe$_{2}$O\emph{M}$_{2}$ \emph{M}\,=\,(S, Se) which are structurally similar to the iron pnictide superconductors.  They are also in close electronic phase diagram proximity to the iron pnictides. Nuclear magnetic resonance (NMR) revealed a critical slowing down of nematic fluctuations as observed by the spin-lattice  relaxation rate ($1/T_1$).  This is complemented by the observation of a change of electrical field gradient over a similar temperature range  using M\"ossbauer spectroscopy. The neutron pair distribution function technique applied to the nuclear structure reveals the presence of local nematic $C_2$ fluctuations over a wide temperature range while neutron diffraction indicates that global $C_{4}$ symmetry is preserved. Theoretical modeling of a geometrically frustrated spin-$1$ Heisenberg model with biquadratic and single-ion anisotropic terms provides the interpretation of magnetic fluctuations in terms of hidden quadrupolar spin fluctuations. Nematicity is closely linked to geometrically frustrated magnetism, which emerges from orbital selectivity. The results highlight orbital order and spin fluctuations in the emergence of nematicity in Fe-based oxychalcogenides. The detection of nematic fluctuation within these Mott insulator expands the group of iron-based materials that show short-range symmetry-breaking.
\end{abstract}

\section{Introduction}
Nematic phases occur in a variety of very different physical systems throughout nature.    A nematic phase is formed when a discrete rotational symmetry is broken or reduced, while the translational symmetry is preserved. Nematicity has been studied mostly in the structural organization of soft matter such as liquid crystals.~\cite{{deGennes_book1974},{DEGENNES1972753},{AdriaanDeVries1985}}  In these materials it is easy to visualize the nematic state's defining characteristic: the presence of a preferred axis along which liquid crystal molecules are, on average, aligned. This spatial anisotropy is generally described by $C_2$ rotational symmetry.  In the context of cuprate high-temperature superconductivity (HTSC) bond-nematicity was introduced as a competing order with HTSC itself.  In recent years, nematicity has become the central focus of efforts to illuminate mechanism(s) of HTSC in iron-based superconductors (FeBS). By now, a variety of iron pnictides have been shown to exhibit electronic nematic states~\cite{0953-2048-25-8-084005,Fernandes_NatPhys2014} and in numerous theoretical pictures, the nematic phase is thought to be intimately related to the superconducting mechanism in iron-based materials.~\cite{Kretzschmar_NaturePhys}

The origin of the nematic phase in iron superconductors is heavily debated.  Two scenarios are proposed:~\cite{Fernandes_NatPhys2014}  One proposal suggests that anisotropic spin fluctuations are necessary precursors of the nematic ordering that has been experimentally observed;~\cite{Kasahara_Nematicity_2012} the other claims that ferro-orbital ordering involving $d_{xz,yz}$ orbitals is responsible for nematicity.  Complicating this debate is the fact that various Fe-based materials exhibit different magnetic order. Nematicity has been mainly studied in the 122 iron pnictides such as BaFe$_{2}$As$_{2}$, but recently nematic behavior was reported in the 1111 material LaAsFeO$_{1-x}$F$_{x}$.~\cite{PSSB:PSSB201600214-LaFeAsO} An important question emerges.  Is nematicity a general phase behavior that exists in other Fe-based materials across the electronic phase diagram?  For example, do Fe-based parent materials that are Mott insulators also exhibit nematicity?  What role does electron correlation in multi-orbital systems play in the formation of the interlinked nematic and spin density wave (SDW) phase and the competing superconducting phase.~\cite{Matsuura_max_Nem}    Increasing evidence points to an electronic mechanism of nematicity which would place the nematic order in the class of correlation-driven electronic instabilities, like superconductivity and density-wave transitions. 

\begin{figure}
	\centering
	\includegraphics[width=6in]{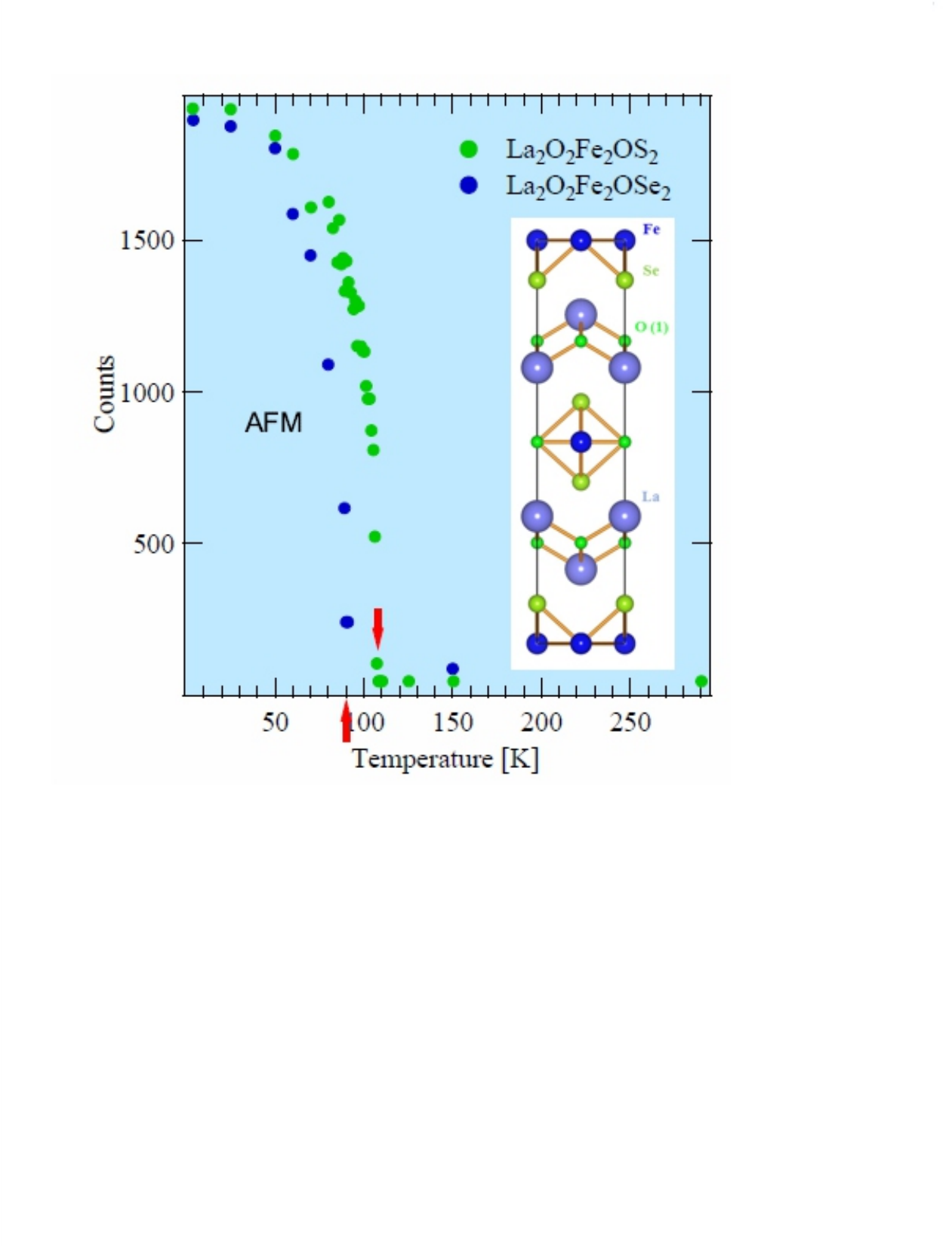}
	\caption{\label{fig:OrderParameter}  The intensity of an antiferrommagnetic peak $\bf{(103)}$ in {\cSSe} as a function of temperature.  At room temperature {\cSSe} are in the paramagnetic phase. The development of magnetic ordering is observed for both materials as the temperature decreases.  The N\'eel temperatures \begin{math}T_{N}\end{math}, indicated by the arrows, for \emph{M} = S and Se are 107.2(6) $\pm$ 0.06\,K and 90.1(9) $\pm$ 0.16\,K, respectively.  The inset shows the crystal structure of \cSe.}
\end{figure}

In this work, we report a combined experimental and theoretical investigation of magnetism intrinsic to the correlation-induced~\cite{Zhu-2010} Mott insulators La$_{2}$O$_{2}$Fe$_{2}$O(S, Se)$_{2}$ (see Figure \ref{fig:OrderParameter}). Neutron powder diffraction was employed to make a magneto-structural comparison of the materials. A short range nematic fluctuating behavior was observed. A critical slowing down of nematic fluctuations observed in the nuclear magnetic resonance (NMR) spin relaxation rate; this result was complemented by M\"ossbauer spectroscopy data. In contrast to Fe-based superconductors, orbital nematicity does not play a role in the iron oxychalcogenides, since a $d_{xz,yz}$ orbital degeneracy is not present due to an alternating orientation of the Fe\emph{M}$_{4}$O$_2$ octahedra. Therefore, a different mechanism must be responsible for the nematicity we observe in {\cSSe}.  Since these are Mott insulators, a strong-coupling  view should be relevant.  Our magnetic neutron diffraction data reveal the establishment of antiferromagnetic (AFM) ordering and the similarity of magnetic behavior in these materials.  Our theoretical modeling explains the nematic fluctuations as being  related to quadrupolar fluctuations in a strongly frustrated spin-1 magnet.  This agrees well with NMR \begin{math}1/T_{1}\end{math} data.  The quadrupolar fluctuations are closely related to geometric frustration in the non-collinear spin structure, which itself arises as a consequence of strong orbital selectivity in the compounds. Thus, orbital-selective Mott correlations are important for understanding magnetism in the Mott-insulating states of the oxychalcogenides. Such selective-Mott correlations have recently been the focus of attention in the iron-selenides~\cite{{PhysRevLett.110.067003},{Si_Rong_2017},{Si_Abrahams_NatRev2016},{0953-8984-28-45-453001}} and the present study provides a new instance for the relevance of orbital-selective Mott physics in Fe-based materials.

\section{Experimental and Theoretical Results}

\subsection{Neutron Diffraction}
Neutron powder diffraction experiments were performed on powder samples of La$_{2}$O$_{2}$Fe$_{2}$O(S, Se)$_{2}$ using the C2 high-resolution diffractometer at the  Canadian Nuclear Laboratories in Chalk River, Ontario. The C2 diffractometer is equipped with an 800 wire position sensitive detector covering a range of 80 degrees. Data were collected in the 2$\theta$ angular range from $5^\circ\mathrm{}$ to $117^\circ\mathrm{}$ using a Si (5 3 1) monochromator at a wavelength $\lambda$ of 1.337 
\text{\AA}. The lattice parameters of {\cSSe} were determined by Rietveld analysis of the diffraction data: for La$_{2}$O$_{2}$Fe$_{2}$OS$_{2}$  [\emph{a} = 4.0453(5) \text{\AA}, \emph{c} = 17.9036(3)
\text{\AA}] and {\cSe}  [\emph{a} = 4.4087(5)
\text{\AA}, \emph{c} = 18.6005(3)
\text{\AA}].  
We did not observe nuclear Bragg peak splittings, that would be indicative of a structural phase transition, in the {\cSSe} diffraction data collected over the temperature range from 4 to 300\ K.  Thus we did not detect a change of lattice symmetry over this temperature range.     

We verified the consistency of the data with the non-collinear 2-$\bf{k}$ magnetic structure  previously obtained for {\cSe}.~\cite{PhysRevB.89.100402} The high-temperature paramagnetic (PM) phase is compared to the 2-$\bf{k}$ antiferromagnetic phase in Fig. \ref{fig:SpinMagnetic}. The Sarah suite of programs~\cite{Wills2000680} was used to analyze the representations and provide the magnetic basis vectors for refinement with Fullprof.~\cite{RODRIGUEZCARVAJAL199355,roisnel2001WinPLOTR} The magnetic cell is commensurate and is doubled, in both \emph{a} and \emph{c}, with respect to the structural cell. The ordering is associated with \begin{math}\bf{k}_1 \end{math} = (1/2, 0, 1/2) and \begin{math}\bf{k}_2 \end{math} = (0, 1/2, 1/2), and the single Fe site on (1/2, 0, 0) in the nuclear \begin{math}I4/mmm\end{math} cell is described by two distinct orbits governing the two (1/2, 0, 0) and (0, 1/2, 0) Fe sites that are independent in the magnetically ordered state. However, we note that for powder samples the diffraction pattern is  indistinguishable from the pattern assigned as the collinear AFM3 model in the literature.~\cite{PhysRevB.81.214433,PhysRevB.89.100402} Neutron diffraction shows that the magnetic structures of {\cSSe} are similar with a distinction in the AFM onset temperatures \begin{math}T_{N}\end{math}. The magnetic peak intensity over a temperature range from 4 to 300\ K, shown in Figure \ref{fig:OrderParameter}, is proportional to the square of the magnetic order parameter.  The magnetic peak intensity data reveals the onset of AFM ordering at 90.1(9) $\pm$ 0.16 K and 107.2(6) $\pm$ 0.06 K for {\cSe} and {\cS}, respectively.  The \begin{math}T_{N}\end{math} value for {\cSe} is in excellent agreement with values in the literature.~\cite{Zhu-2010,PhysRevB.81.214433,PhysRevB.89.100402}

\subsection{Nuclear Pair Distribution Function}
Nuclear pair distribution function (PDF) measurements on {\cSSe} were performed using the Nanoscale Ordered Materials Diffractometer (NOMAD)~\cite{NEUEFEIND201268} beamline at the Spallation Neutron Source (SNS) at Oak Ridge National Laboratory (ORNL). PDF measurements are well-suited to provide details of the structural arrangement of atoms. Pair distribution function analysis involves obtaining the Fourier transform of the measured total scattering intensity \begin{math}S(Q)\end{math} in order to obtain a real space representation of inter-atomic correlations.  The pair distribution function is given as \begin{math}G(r) = {\frac{2}{\pi}}\int_{0}^{Q_{max}}Q[S(Q)-1]Sin(Qr)dQ\end{math} where \begin{math}Q\end{math} is the scattering vector and \begin{math}r\end{math} is the interatomic distance. PDF is a total scattering method \emph{i.e.}, meaning that both Bragg and diffuse scattering intensity data is simultaneously collected.  Therefore, the technique is sensitive to deviations from the average structure.~\cite{citeulike:13633118} Our PDF data were generated from the total scattering experiments using neutrons of a wavelength band from 0.1 to 2.9 \text{\AA}. The maximum momentum transfer used for the Fourier transform was 31.4 \text{\AA}$^{-1}$. The instrument parameters Q$_{damp}$ = 0.017 \text{\AA}$^{-1}$ and Q$_{broad}$ = 0.019 \text{\AA}$^{-1}$ were fixed for all of our fits and structural refinements of the data were performed using the \texttt{PDFGUI}~\cite{Farrow_2007}  program and \texttt{DIFFPY-CMI}~\cite{Juhas:po5132} suite. 

\begin{figure}
	\centering
	\includegraphics[width=6in]{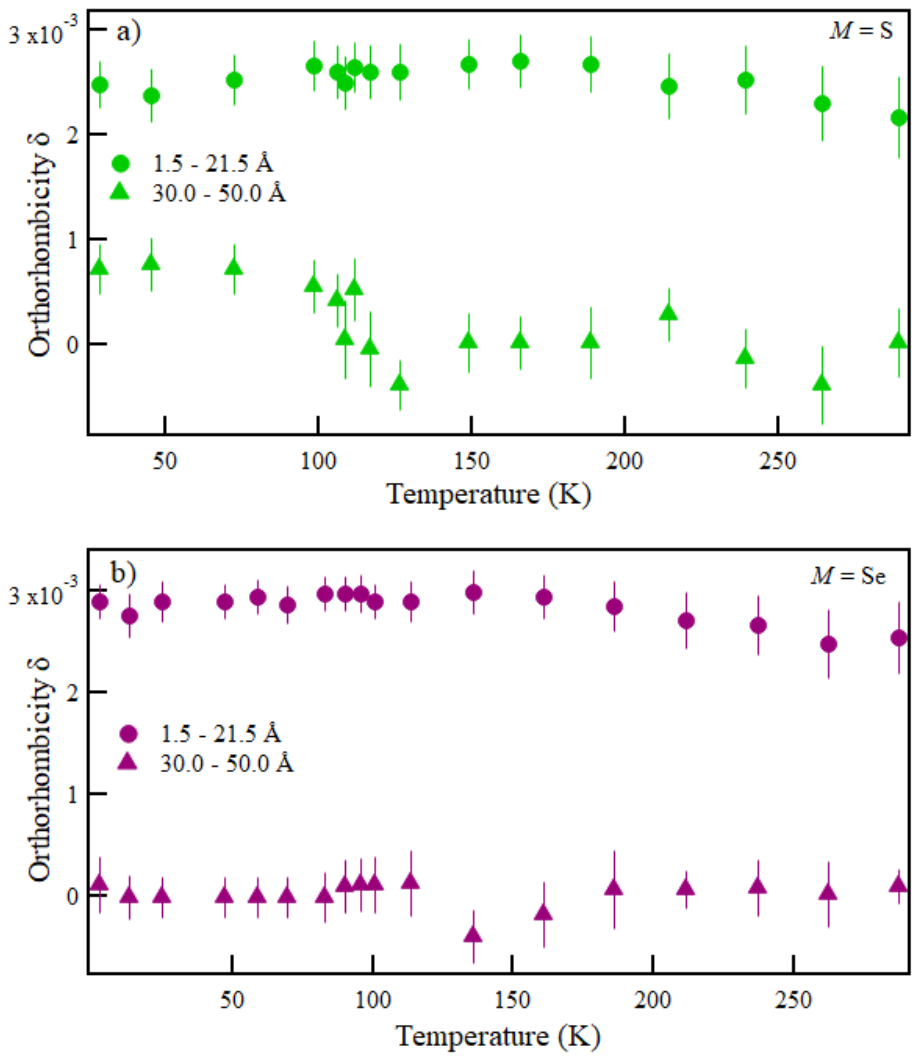}
	\caption{\label{fig:Orthorhombicity}  Temperature dependence of orthorhombicity for short-fitting range (1.5-21.5) \text{\AA} and long-fitting range (30.0-50.0) \text{\AA} for a) \emph{M} = S and b) \emph{M} = Se. In both of these figure orthorhombicity is along y-axis and temperature is in \emph{x}-axis. A solid circle symbol indicates the short-fitting range (1.5-21.5) \text{\AA} and a solid triangle symbol indicates the long-fitting range.}
\end{figure}

 We performed data refinements using an orthorhombic model of {\cSSe} with a sliding fit range from (1.5-21.5) \text{\AA} to (29.5-49.5) \text{\AA} in 1 \text{\AA} steps to investigate the temperature dependent structure on these various length scales. This resulted in 29 fits for each temperature at which PDF data were collected. We have extracted the orthorhombicity parameter $\delta$ = \begin{math}(a-b)/(a+b)\end{math} for both length scale and temperature series sequential refinements where \begin{math}a, b\end{math} are lattice parameters. Figure \ref{fig:Orthorhombicity} shows the refined orthorhombicity $\delta$ extracted from fits of neutron PDF data.  Structural models obtained from refining PDF data are strictly valid~\cite{proffen2005a} within the refined length scale; this specificity condition enables the probing of local structure on different length scales by varying the refinement range.  By conducting such fittings, we observed higher orthorhombicity for short-range fitting (1.5 - 21.5) \text{\AA} compared to long-range fitting (30.0-50.0) \text{\AA}. This suggests that by tracking the orthorhombicity parameter we observe local scale distortions from tetragonal to orthorhombic structure. These spatially limited distortions represent fluctuating nematic order~\cite{PhysRevLett.119.187001, PhysRevB.98.180505} over a broad range of temperatures. It should be emphasized that neutron powder diffraction, from the same samples, provides clear evidence that the average, long-range structures of {\cSSe} remain tetragonal throughout the high and low temperature regimes.~\cite{PhysRevB.99.024109} However, the PDF results provide evidence of local symmetry breaking that results in nematic \begin{math}C_2\end{math} regions of  fluctuating order. \cite{PhysRevMaterials.1.074412, PhysRevB.94.094102, PhysRevB.98.180505, PhysRevLett.119.187001}

\subsection{M\"ossbauer spectroscopy}

In order to investigate the magnitude and the orientation of the ordered static Fe magnetic moment with respect to the electric field gradient (EFG) principal axis within the distorted FeO$_2$S$_4$ octahedra, and to study the modification of the electronic surroundings via the electric quadrupole interaction, $^{57}$Fe M\"ossbauer spectroscopy was performed on {\cS} and is discussed in comparison to similar work on {\cSe}.~\cite{{PhysRevB.90.184408-Marco-gunter}}  The $^{57}$Fe nucleus probes the onsite magnetism of the Fe ion, therefore any significant changes in orbital and spin degrees of freedom should be reflected in the $^{57}$Fe M\"ossbauer spectra.

Figure \ref{fig:NewFig2_MossVzz}a shows the low temperature zero field $^{57}$Fe M\"ossbauer spectra for \cSe\ and \cS. At high temperatures, in the paramagnetic (PM) state, we observe a single asymmetric doublet (spectra not shown here) with a quadrupole splitting of $\Delta$ = 1.95 mm/s corresponding to a  principal EFG component of $V_{\mathrm{zz}}$\,=117 V/$\mathrm{\text{\AA}}^2$. The basic features of the spectra in the PM phase look similar to previously reported {\cSe} results.~\cite{{0953-8984-22-34-346003}, {doi:10.1021/ja711139g},{PhysRevB.90.184408-Marco-gunter}} As the temperatures goes below 110\,K a sextet develops thereby indicating the long range magnetic ordering of \cS.  

M\"ossbauer spectra were analyzed by using the full static Hamiltonian~\cite{}

\begin{align}
    H_s = &\frac{eQ_{\mathrm{zz}}V_{\mathrm{zz}}}{4I(2I-1)}\left[(3I_z^2-I^2)+\frac{\eta}{2}(I_+^2+I_-^2)\right] \notag\\
                &-g_I\mu_NB\left(\frac{I_+e^{-i\phi}+I_-e^{+i\phi}}{2}\sin\theta+I_z\cos\theta\right) \;,
\end{align}
with nuclear spin operators $I_z$, $I_{+}\,=\,I_x+iI_y$, and $I_{-}\,=\,I_x-iI_y$. Here $\mathbf{B}$ is the hyperfine field at the $^{57}$Fe site while $Q$, $g_\mathrm{I}$, and $\mu_\mathrm{N}$ indicate the nuclear quadrupole moment, g factor and magneton, respectively. The polar angle $\theta$ and the azimuthal angle $\phi$ describe the orientation of the Fe hyperfine field $\mathbf{B}$ with respect to the EFG $z$ axis. The azimuthal angle $\phi$ was set to zero to avoid cross correlation. This is justified, because the asymmetry parameter $\eta$ is small and the polar angle $\theta$ is close to zero for both systems. For both samples, the whole temperature series was fitted simultaneously to determine $\eta$, then $\eta$ was fixed, and then the spectra were fitted individually. Within error bars, both samples {\cSSe} have $\eta\approx 0.1$.

In the magnetically ordered state a clear magnetic sextet is observed in \cSSe\ data, indicating the presence of a long range ordered (LRO) state. The nature of this LRO state is the same for both systems in terms of the magnitude of the ordered moment (magnetic hyperfine field $B_{\mathrm{hyp}}(T)$) and its temperature dependence. The lines in Figure \ref{fig:NewFig2_MossVzz}b) indicate fits of the sub-lattice magnetization $M(t)$ by using the equation $M(t)\propto B_{\mathrm{hyp}}(T)\propto (1-T/T_N)^{\beta_\mathrm{crit}}$ applied to the magnetic hyperfine field data.  The magnetic critical exponents ${\beta_\mathrm{crit}}$ can be estimated as a fit parameter (see Table~\ref{moess-properties}).

The $^{57}$Fe M\"ossbauer spectroscopy provides clear evidence that in both systems the angle between the strongest EFG component ( local $z$ main axis) and the magnetic hyperfine field $B$, $\theta$, is equal to 0 in the magnetically ordered regime, \emph{i.e.}, the magnetic hyperfine field $\mathbf{B}$ is oriented parallel to the local $z$ axis of the EFG principal axis system. The strongest EFG component is aligned along the O-Fe-O chains in agreement with local spin density approximation + \emph{U} (LSDA+ \emph{U}) calculations.~\cite{0953-8984-22-34-346003} This indicates that the Fe moments are oriented parallel to O-Fe-O chains. Therefore, both \cS\ and \cSe\ exhibit non-collinear magnetic order with a 90 degree angle between different magnetic sublattices. This is consistent with three different spin structures discussed for \cSe\. ~\cite{PhysRevB.90.184408-Marco-gunter} However, the  $^{139}$La NMR experiments presented below decisively identify the 2-$\bf{k}$ structure as the actual magnetic structure in both systems.

The temperature dependence of the EFG is comparable in {\cSSe}. At $\approx 1.4\,T_N$ the principal component $V_{zz}$ decreases significantly in a near-linear fashion (Figure \ref{fig:NewFig2_MossVzz}c).  
Previously reported thermal expansion data of \cSe\ also showed an unusual change in this temperature range.~\cite{PhysRevB.81.214433} It is known~\cite{PhysRev.133.A787} that thermal behavior can influence the values of the EFG tensors; however, another relevant possibility is that any degree of orbital electronic anisotropy can significantly modify the orbital occupancy. This would lead to a change in the valence contribution reflected in the EFG tensors. This particular scenario is well supported by the NMR spin-lattice relaxation rate data presented below. 

\subsection{Nuclear Magnetic Resonance (NMR)}

Figure~\ref{fig:NewFig3_LaNMRSweep} shows the representative $^{139}$La field sweep NMR spectra on \cS. Each spectrum consists of three pairs of satellites, and one split central transition. For lower symmetries, such as tetragonal systems, in the presence of finite EFG at the probed nucleus with nuclear spin $I=7/2$, one would expect such NMR spectra. Upon lowering $T$, the spectral shape remains unchanged down to 109\,K (the line in the temperature range 109-105\,K could not be resolved). This indicates the AFM ordering within the system. In the ordered state the spectral shape changes considerably because of the development of the static internal magnetic field associated with Fe ordered moment. 
The field swept NMR spectra can be well described by diagonalizing the full nuclear Hamiltonian
\begin{align}\label{nuclearHamiltonian}
	H = &h\gamma \left[(\mathbf{B} +\mathbf{B}_{\parallel} + \mathbf{B}_{\perp})I + h \frac{\nu}{6}(3I_z^2-I^2 + \eta (I_x^2+I_y^2)\right],
\end{align}
where $\mathbf{B}$ is the external magnetic field, $z$ axis being the principle axis of EFG and $\frac{\gamma}{2\pi}=$ 6.0146 $\frac{MHz}{T}$. $\mathbf{B}_{\perp}$ ($\mathbf{B}_{\parallel}$) are the hyperfine fields (internal fields) at the La site due to the Fe ordered moment. At temperatures $T>T_N$ NMR field sweep spectra can be adequately described without any additional internal fields ($\mathbf{B}_{\perp}$ = 0\,mT, $\mathbf{B}_{\parallel}$ = 0\,mT); however two different local fields $\mathbf{B_1}$ and $\mathbf{B_2}$ at magnetically nonequivalent $^{139}$La sites must be introduced in Eq. (\ref{nuclearHamiltonian}) to describe the observed spectra in the ordered state. The shape of the presented NMR spectra look very similar to the Se variant La$_2$O$_2$Fe$_2$OSe$_2$,~\cite{PhysRevB.90.184408-Marco-gunter} and we ascertain a non-collinear magnetic structure for La$_2$O$_2$Fe$_2$OS$_2$ in agreement with the $^{57}$Fe M\"ossbauer results. The presence of two different sub-spectra of equal intensity with local hyperfine fields parallel and perpendicular to the $c$-axis unambiguously confirm the 2-$\bf{k}$ magnetic structure to be realized in both systems.~\cite{PhysRevB.90.184408-Marco-gunter}  The applied fields do not change the magnetic structure of the sublattices due to anisotropy fields much larger than 7\,T. This was checked by $^{57}$Fe M\"ossbauer spectroscopy, where a model based on pure superposition of internal and external field reproduces the measured spectra, similar to the NMR results.

There are no $^{139}$La NMR spectral anomalies near or below 150\,K in {\cS} nor near or below 125 K in {\cSe}.~\cite{PhysRevB.90.184408-Marco-gunter}  In contrast, there are almost linear decreases in the principal component $V_{zz}$ of EFG near 125\,K and 150\,K in {\cSe} and {\cS}, respectively. The changes in the $^{57}$Fe M\"ossbauer EFG suggest that the behavior is orbitally-dependent on Fe $3d$-states, because the anisotropy caused by different occupation of the  $d$ orbitals should affect the $V_{zz}$. That such an effect is not discerned in the NMR static parameter is not surprising because the La nucleus is coupled by hyperfine interaction, which is not necessarily sensitive to the on-site Fe magnetism. 

Figure \ref{fig:NewFig4_1_T1} shows the temperature dependence of {\cS} $^{139}(1/T_1)$ data collected at 5.0\,T. The data is qualitatively similar to \cSe\ both in the paramagnetic and antiferromagnetic states.~\cite{PhysRevB.90.184408-Marco-gunter} Upon lowering $T$ below approximately 150\,K ($T^*_{\text{S}}$) the {\cS} $^{139}(1/T_1)$ data  starts to increase reaching a divergence around 108\,K. Similarly, the $^{139}(1/T_1)$ signal is enhanced below approximately 125\,K ($T^*_{\text{Se}}$) and approaches a divergence near 90\,K.   The divergences are associated with the critical slowing down of the Fe spin fluctuations. The enhancements of $^{139}1/(TT_1)$ below ($T^*_{\text{S}}$) and ($T^*_{\text{Se}}$) occur in the same temperature ranges where M\"ossbauer spectra exhibit a decrease of the principal component $V_{\mathrm{zz}}$ of the EFG for {\cSSe}. The enhancements are related to short range correlations above $T_{N}$ which influence the spin-lattice relaxation rate.  We emphasize that the thermal behavior of the $c$-lattice constant of {\cSSe} have been reported to exhibit anomalous contractions near ($T^*_{\text{S}}$) and ($T^*_{\text{Se}}$). The $c$-lattice reductions manifest as a deviation from the Einstein temperature model.~\cite{PhysRevB.81.214433} The lattice change is concomitant with the decrease in $V_{\mathrm{zz}}$ which suggests that EFG is influenced by the change in the crystal field associated with octahedral height. 

Generally, the NMR spin-lattice relaxation rate is proportional to the imaginary part of the  dynamic susceptibility at the Larmor frequency ($1/T_1T \sim \chi^{``}(\textbf{q},\omega_0)$), where $\textbf{q}$ is the wave vector and  $\omega_0$ is the Larmor frequency.~\cite{{Moriya-doi:10.1143/JPSJ.18.516},{smerald-PhysRevB.84.184437}} Therefore NMR $^{139}(1/T_1)$ is very sensitive to fluctuations related to the orbital and spin degrees of freedom at the Fe site. The NMR $^{139}(1/T_1)$ data in Figure \ref{fig:NewFig4_1_T1} provide a clear indication of a continuous slowing down of the Fe spin or orbital degrees of freedom far from the AFM ordered phase.  
Similar observations were also reported for LaFeAsO~\cite{{PhysRevLett.109.247001-Imai},{doi:10.1143/JPSJ.77.073701-nakai}, {PSSB:PSSB201600214-LaFeAsO}} where, just below the structural transition, $^{139}(T_1)$ starts to increase before it diverges at the antiferromagnetic ordering temperature. Both {\cSSe} fit well into this framework albeit they have not been observed to undergo structural phase transitions.    

For $T<T_N$, the spin-lattice relaxation rate $^{139}(1/T_1)$ is strongly reduced over a very short temperature interval. Below 87\,K, for La$_2$O$_2$Fe$_2$OS$_2$ $^{139}(1/T_1)$ can be fitted by activation gap-like behavior $\sim T^2\exp(-\Delta/T)$, with a gap $\Delta =$70\,K, where as for the Se variant it is $\Delta =$ 55\,K.  This observation is consistent with opening up of a spin gap as reported in a neutron scattering study.~\cite{Lumsden-jpcm-2010} For $T<30$~K, we find a smooth crossover from gap-like behavior to a gapless regime, where $1/T_{1}\simeq T^{3}$. Therefore, two distinct regimes are observed where
$1/T_{1}\simeq T^{3}$ for $T<<\Delta$, while $1/T_{1}\sim T^2\exp(-\Delta/T)$ for $T>\Delta$. 

The NMR $1/T_{1}$ behavior for 30\,K $<$ $T$ follow an unusual $T^{3}$ power-law form  
which suggests the activation of an additional spin fluctuation channel.  We now address the origin of this behavior within the framework of an $S$ = 1 Heisenberg model with appreciable geometric frustration, an additional bi-quadratic term, and single-ion anisotropy as is appropriate for Fe-based systems.  

\subsection{Theory}
Motivated by earlier local density approximation + dynamical mean-field theory (LDA+DMFT) calculations in which we estimated an $S=1$ on each Fe site, we begin with the spin $S=1$ bilinear-bi-quadratic model of the square lattice.  This model has been widely employed in context of magnetism and nematicity in Fe-based SCs.~\cite{{PhysRevLett.101.076401}, {Si_Nevidomskyy_arXiv2016},{Note_on_Spin1_2}}  The Hamiltonian is
\begin{equation}\label{Hij}
H=\sum_{i,j}[J_{ij}{\bf S}_{i}.{\bf S}_{j} + K({\bf S}_{i}{\bf S}_{j})^{2} - D(S_{i}^{z})^{2}] \;,
\end{equation}
where ${\bf S}_{i}$ is a spin-$1$ operator on site $i$.  $J_{ij}=J_{1a}$, $J_{ij}=J_{1b}$ such that $J_{1a}$ is ferromagnetic (FM)  and $J_{1b}$ is AFM. $J_{ij}=J_{2}$ for $[ij]$ along the square-diagonal, and $K$ is the coefficient of the biquadratic interaction for $S=1$ along with  $D$ being the single-ion anisotropy term that fixes the orientation of the spin. It is known that this model admits various ordered phases, depending upon the relative values of the $J_{ij}$ and $K$ for a given $D$:~\cite{PhysRevLett.115.116401} a ${\bf q}=(\pi,0)$ collinear antiferromagnetic (CAF) phase, a ${\bf q}=(\pi/2,\pi)$ AFM$^{*}$ phase, a ${\bf q}=(\pi,\pi)$ N\'eel AFM phase, and a ${\bf q}=(\pi,0)$ anti-ferro-quadrupolar (AFQ) phase.  Additionally, the strong geometrical frustration (supported by small value of the ordered moment relative to density functional theory estimates) inherent in the model, along with sizable biquadratic and single-ion terms can yield a non-collinear AF phase as observed in our systems.  The FM and AFM couplings, along with sizable $D$, also favor $D=2$ Ising-like spin fluctuations (as extracted from the critical exponent for $M(T)$). With this in mind,
we have estimated the temperature dependence of the NMR relaxation rate, 
$1/T_{1}\simeq \lim_{\omega\rightarrow 0}\sum_{\bf q}$Im$\frac{\chi({\bf q},\omega)}{\omega}$
by computing the local spin correlation function, $\chi_{ii}(\omega)=\int_{0}^{\infty}dt e^{i\omega t}\langle T[S_{i}^{+}(t)S_{i}^{-}(0)]\rangle$ using quantum-Monte Carlo (QMC) methods applied to the $H$ [Eq.~(\ref{Hij})] above.~\cite{ALPS_project}  The $1/T_{1}$ QMC results (green squares) are shown in Figure \ref{fig:theory_NMR}.  

Within the temperature range $30 < T <T_{N}(\sim100$~K), the experimental data for NMR $1/T_{1}$ fits $AT^{2}$exp$(-\Delta/T)$ to a very good approximation, with $\Delta$ estimated to be approximately 100 K. Moreover, at low temperatures $T<30$~K, we also find that the $1/T_{1}\simeq T^{3}$, indicating gapless spin excitations in the spin fluctuation spectrum.  This is consistent with earlier findings on the Se compound.~\cite{PhysRevB.90.184408-Marco-gunter} Our theoretical findings are consistent with strong suppression of magnetic fluctuations below $T_{N}$, a crossover to a gapped regime of spin fluctuations with a crossover scale $O(50$~K), consistent with available inelastic neutron scattering data, and at low $T<30$~K, another spin relaxation channel opens up, indicating low-energy spin excitations below the gap.

These findings can be understood in by considering that a field-theoretic analysis for a collinear AF yields $1/T_{1}\simeq T^{3}$ for $T>>\Delta$, but $1/T_{1}\simeq T^{2}\Delta$exp$(-\Delta/T)$ for $T<<\Delta$.  For an AFQ ordered state the same analysis yields $T_{1}^{-1}\simeq T^{3}e^{-\Delta/T}$ for $T<<\Delta$, but $T_{1}^{-1}\simeq T^{7}$ for $T>>\Delta$. Both of these are inconsistent with our experimental findings and, at first glance, would rule out both collinear AFM and AFQ phases at low $T$ in the oxychalcogenides.  However, our results are not in conflict with the previously proposed non-collinear AFM phase in the iron oxychalcogenides.~\cite{PhysRevB.89.100402,PhysRevB.90.184408-Marco-gunter}  
Three-magnon processes can be ruled out, since no $T^{5}$ - dependence is observed in our experimental data (nor in the calculations) at any studied temperature.~\cite{PhysRev.166.359}  In considering all of the presented data and our theoretical results, the clear scenario emerges in which the {\cSSe} Mott insulators exhibit a non-collinear magnetic structure that might support the presence of AFQ phases.  Furthermore, these materials each contain gapless excitations in the spin fluctuation spectrum.  An important caveat for the low-temperature region $T<<\Delta$ is the our uncertainty regarding the deviation of NMR $1/T_{1}$ experimental data. We mention that we are not able to rule out the possibility of  impurities or defects causing $1/T_{1}$ signal deviation from gap-like behavior at low temperatures.  The presence of slight impurities was suggested to be the source of the low-temperature upturn in magnetic susceptibility data of {\cSSe}.~\cite{PhysRevB.99.024109} 

Were it possible to electron-dope the iron oxychalcogenides by chemical substitution or strain in order to modify their Mott insulating electronic structure, one would expect the small density of metallic carriers near the Mott transition to strongly scatter off these low-lying spin correlations.  This would involve carriers propagating by emitting and absorbing two magnons in the usual view of a doped Mott antiferromagnet.  The doped carriers would propagate by scattering off dynamical spin fluctuations that are encoded in the one-spin-flip fluctuation spectrum. We note that these processes might give rise to anomalous metallic features but they remain un-investigated.  In addition, the orbital selectivity that arises from the multi-orbital character of the iron oxychalcogenides will also generally lead to incoherent metallicity.  How the interplay between these distinct electronic mechanisms might conspire to generate unconventional ordered state(s) in Mott systems remains a fascinating open issue.  The current work should encourage efforts along these lines.

In summary, we have performed a combined theoretical and experimental study of the iron oxychalcogenides \cSSe.  A 2-$\bf{k}$ non-collinear magnetic structure  was confirmed in both systems. Striking signatures of local nematic fluctuations  in the magnetic responses of these materials as the AFM phase is approached from high temperatures.  Nematic fluctuations are observed to occur by virtue of short lengthscale rotational symmetry breaking.  The neamtic fluctuation data is consistent across a range of different experimental techniques.  Quadrupolar behavior is responsible for the critical slowing down of spin fluctuations was observed in the $^{139}(T_1)$ NMR data.  These emergent quadrupolar fluctuations arise from sizable geometric frustration, which is itself a consequence of strong orbital selectivity of the  compounds.~\cite{PhysRevB.89.024509}  Our theoretical modeling shows that such orbital selectivity underpins numerous behaviors in the correlation-induced Mott-insulating states of the oxychalcogenides.  Along with earlier DMFT work, the present results also establish the relevance of a strong correlation-based approach for iron oxychalcogenides.  Finally, we note that nematicity has been invoked~\cite{Kivelson_Nature1998} in both experimental and theoretical contexts to describe order competing with high-$T_c$ superconductivity in the cuprates and the iron pnictides. Therefore, the current work expands a current topic in iron-based high-temperature superconductivity while demonstrating the underlying relevance of structural nematicity to larger set of quantum materials.

    
\newpage{}
\vspace{-70mm}
\begin{flushleft}
	{\large \bf{Figure 1}}
\end{flushleft}
\clearpage{}

\newpage{}
\vspace{-20mm}
\begin{flushleft}
	{\large \bf{Figure 2}}
\end{flushleft}
\clearpage{}

\newpage{}
\begin{figure}
	\centering
	\includegraphics[width=6in]{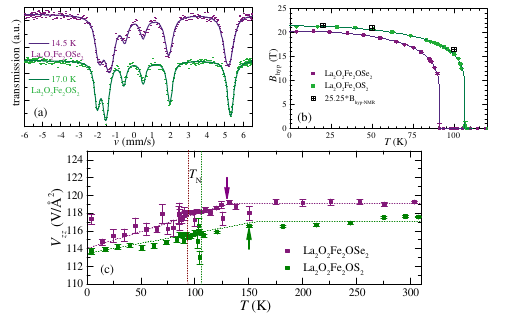}
	\caption{\label{fig:NewFig2_MossVzz}{{\bf{a)}} Low temperature $^{57}$Fe M\"ossbauer spectra of \cSe\ ~\cite{PhysRevB.90.184408-Marco-gunter} and \cS\ fitted with the full static Hamiltonian as described in the text. A single sextet indicates full long-range magnetic order rapidly arising below $T_N$. {\bf{b)}} Magnetic hyperfine field $B_{hyp}$ or the magnetic order parameter determined from $^{57}$Fe M\"ossbauer measurements.  The lines denote the fit of sub-lattice magnetization by using the equation $M(t)\propto B_{\mathrm{hyp}}(T)\propto (1-T/T_N)^{\beta_\mathrm{crit}}$. {\bf{c)}} The temperature dependence of the principal axis component $V_{zz}$ of the EFG decreases above $T_N$ for \cSe\ and \cS. Dotted lines are the guides to the eye.}
	\label{fig:NewFig2_MossVzz}}
\end{figure}
\begin{flushleft}
    {\large \bf{Figure 3}}
\end{flushleft}
\clearpage{}

\newpage{}
\begin{figure}
	\centering
	\includegraphics[scale =0.5]{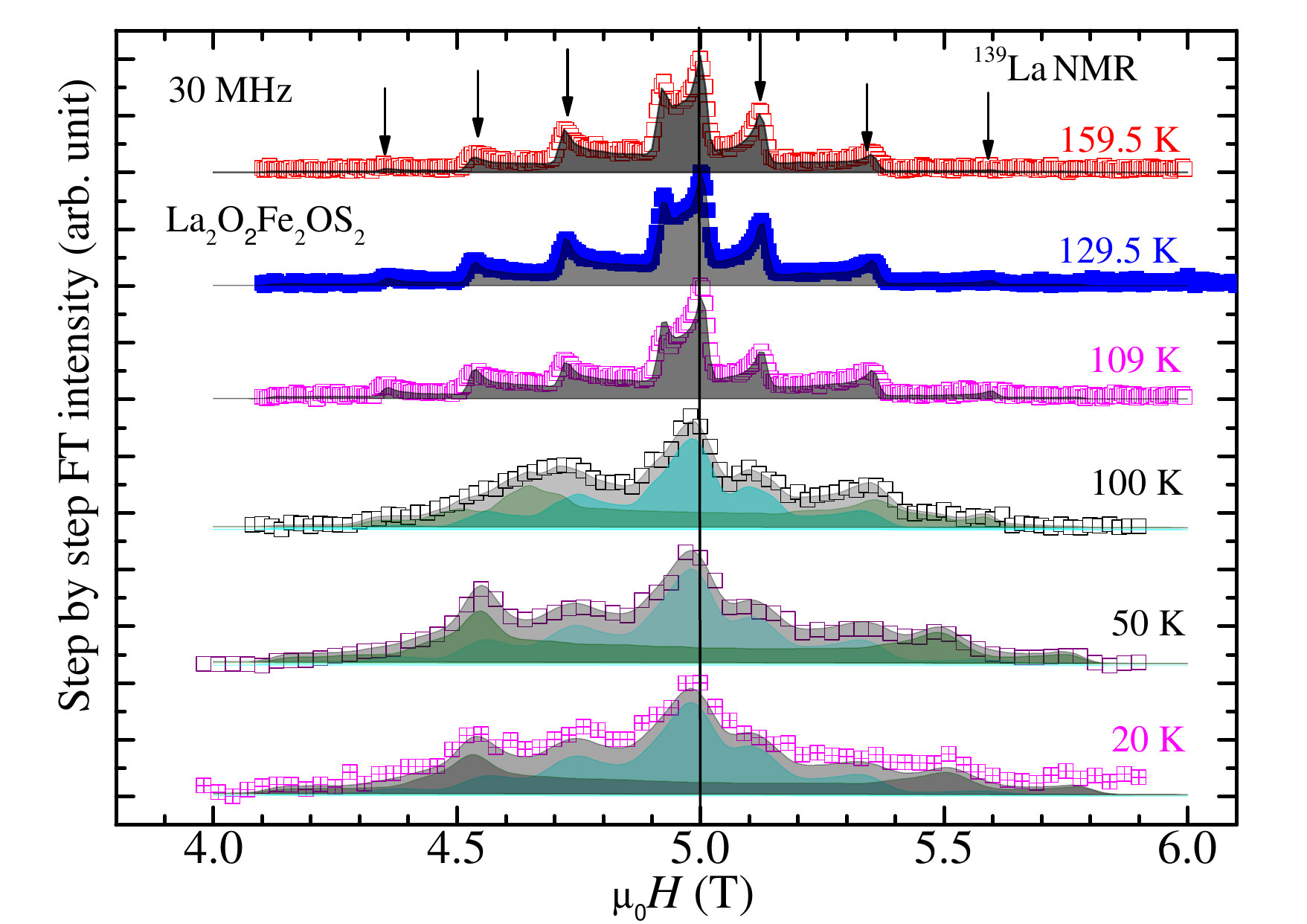}
	\caption{\label{fig:NewFig3_LaNMRSweep} The $^{139}$La field sweep NMR spectra on \cS taken at 30\,MHz. In general, for a nuclear spin ($I = 7/2$), in a lower symmetry system such as orthorhombic and tetragonal, for the presence of finite EFG at the La site, there are seven peaks due to the selective transitions between the levels (m m-1, where m = 7/2, 5/2,... and -5/2). For \begin{math} T > T_{N}\end{math} typical non-magnetic $I = 7/2$ powder spectra are observed. Arrows indicate the corresponding satellite transitions.  Black solid lines are simulated spectra. Green and red solid lines at 100\,K, 50\,K and 20\,K represent calculated sub spectra due to magnetically nonequivalent sites. The vertical line indicates the field at which $^{139}($\begin{math}1/T_{1})\end{math} were measured.}
\end{figure}
\begin{flushleft}
     {\large \bf{Figure 4}}
\end{flushleft}
\clearpage{}

\newpage{}
\begin{figure}
	\center
    \includegraphics[scale=0.5]{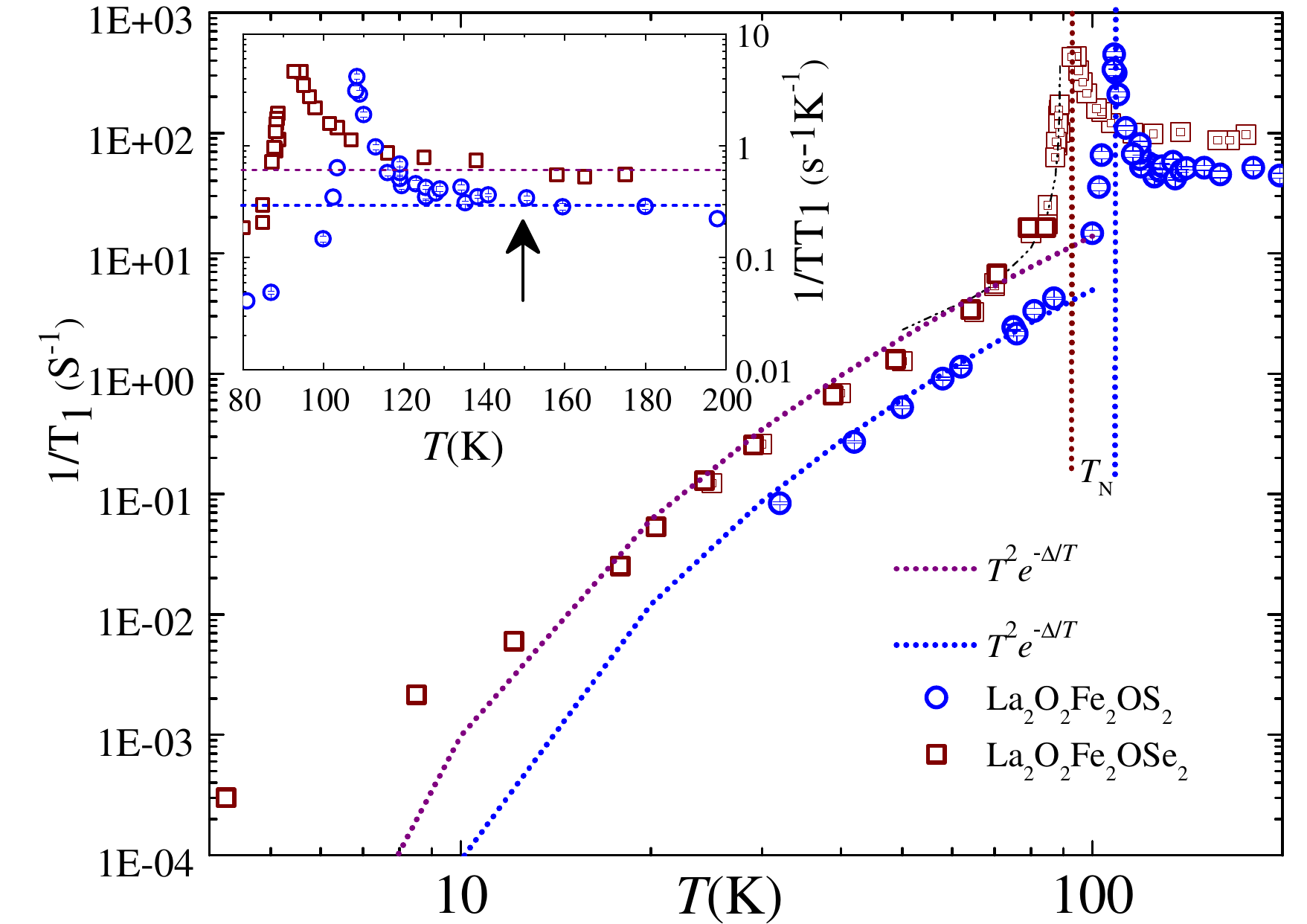} 
	\caption{ \label{fig:NewFig4_1_T1} Temperature dependence of the $^{139}$La spin-lattice relaxation 	rate of \cS\ (filled circle) and \cSe (filled square)
    ~\cite{PhysRevB.90.184408-Marco-gunter}. Lines indicate the activation gap model $\sim T^2\exp(-\Delta/T)$. Dotted vertical lines denote the AFM ordering temperature. Inset shows the $1/T_1T$ versus temperature, while arrow marks the temperature regime where principal component $V_{zz}$ of the EFG start to decrease.}
\end{figure}
\begin{flushleft}
	{\large \bf{Figure 5}}
\end{flushleft}
\clearpage{}

\newpage{}
\begin{figure}
	\center
	\includegraphics[width=5in]{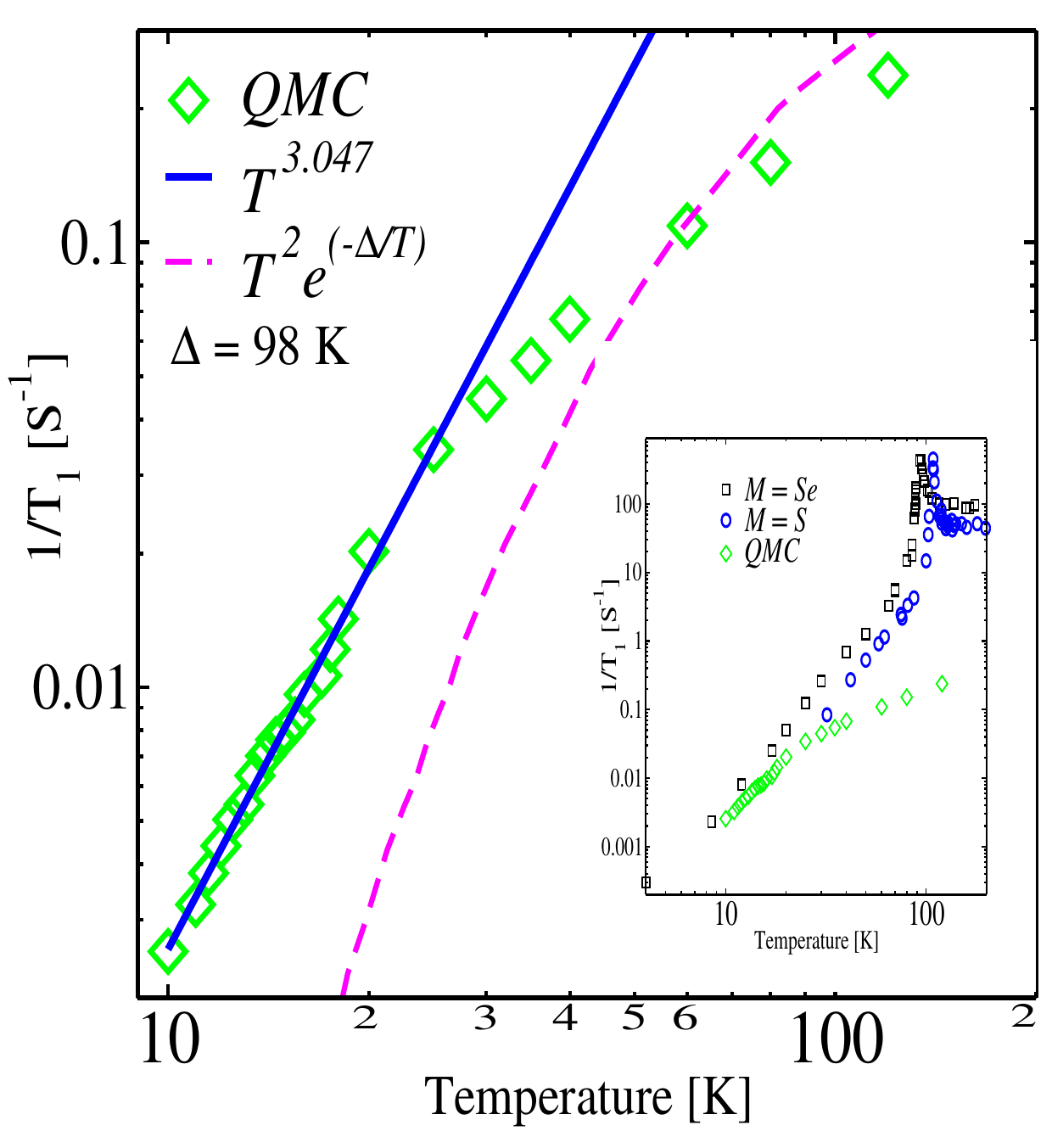}
	\caption{\label{fig:theory_NMR} The nuclear magnetic resonance (NMR) spin relaxation rate 
($1/T_{1}$) as a function of temperature $T$ for {\cSe} studied in this work.  A large-lattice size Quantum Monte Carlo technique was used by employing the Algorithms and Libraries for Physics Simulations (ALPS) to estimate ($1/T_{1}$). By doing so, we computed the local transverse spin-spin correlation function for a $S=1$ Heisenberg model with geometrical frustration, biquadratic exchange, and single-ion anisotropy as relevant for the Fe-based systems.  The QMC results (green squares) were fitted with $T^{2}\exp(-\Delta/T)$ (magenta dashed line) and $T^3$ (blue line) in the high-T and low-T regions, respectively. The main panel indicates that a $1/T_{1} 
\simeq T^{2}\exp(-\Delta/T)$ form for $T>40$~K, with an estimated spin gap of 
98 K, smoothly evolves into $1/T_{1}\simeq T^{3}$ for $T<30$~K.  The low-$T$ power law behavior emerges due to an additional two-magnon relaxation channel opening up at low $T$.  Good accord between theory and  experimental NMR data can be seen (see inset). }
\end{figure}
\vspace{-5mm}
\begin{flushleft}
	{\large \bf{Figure 6}}
\end{flushleft}
\clearpage{}
\newpage{}

\begin{figure}
\center
\includegraphics[width=6in]{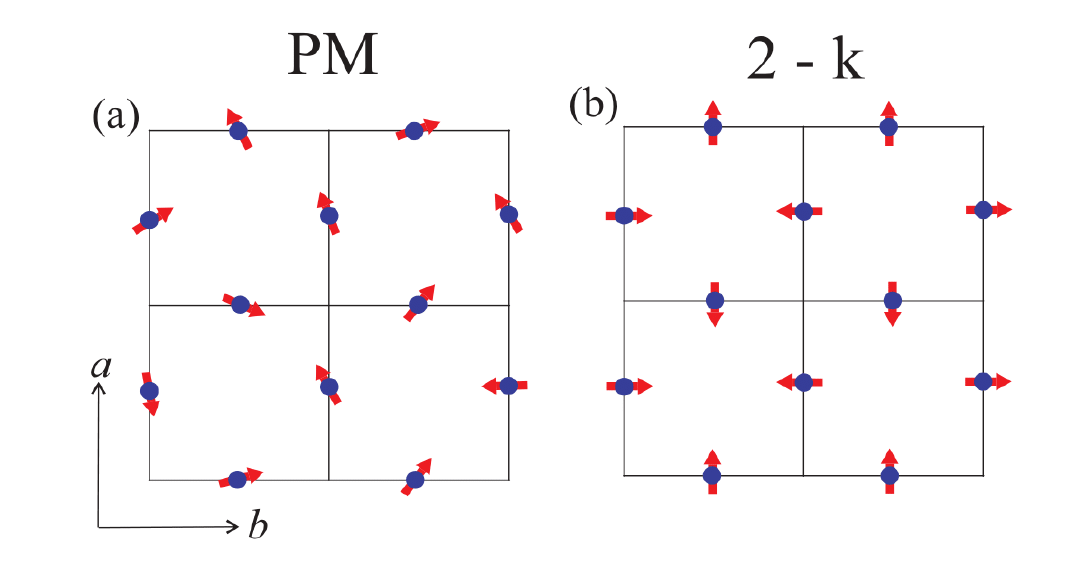}
\caption{\label{fig:SpinMagnetic}Schematic illustrations of magnetic ordering.  For spins (red arrows) located at Fe atom positions (blue circles), the arrangements are given for (a) the paramagnetic (PM) high-temperature phase.  The low-temperature antiferromagnetic phase (b) is the non-collinear 2 - $\bf{k}$ arrangement.
}
\end{figure}
\vspace{-5mm}
\begin{flushleft}
{\large \bf{Figure 7}}
\end{flushleft}
\clearpage{}
\newpage{}


\begin{table}
	\begin{tabular}{lrrr}\hline
		property&\cSe&\cS& \\\hline\hline
		$T_N$ (K)&90.3(2)&106.7(3)\\
		$B_{hyp}(0)$ (T)&20.32(2)&21.43(1)\\
		$\beta_{\mathrm{crit}}$&0.143(1)&0.140(1)\\  
		$\Delta_0$ (mm/s)&1.069(1)&1.044(1)\\
		$\Theta_D$ (K)&300(7)&298(4)\\\hline
	\end{tabular}
	\caption{Properties of the measured oxychalcogenide compounds obtained by  M\"ossbauer spectroscopy. The AFM transition temperature T$_N$, low-temperature $^{57}$Fe saturation hyperfine field $B_\mathrm{sat}$, and the critical exponent $\beta_{\mathrm{crit}}$ are found to vary only in small ranges. $\Theta_D$ is the Debye temperature.}
	\label{moess-properties}
\end{table}


\begin{addendum}
 \item R. Sarkar and H-H. Klauss are partially supported by the
DFG through SFB 1143 for the project C02. VG was partially supported by the DFG through GR 4667.
L.C.'s work is presently supported by CNPq (Grant No. 304035/2017-3). 
Acknowledgment (L.C.) is also made to the Physical Chemistry Department at Technical University Dresden for hospitality.  ZJU work  was supported by the National Basic Research Program of China under grant No. 2016YFA0300402 and 2015CB921004, the Nature Science Foundation of China (Grant No. 11374261 and 11974095), and the Fundamental Research Funds for the Central Universities of China.  Work at Lawrence Berkeley National Laboratory was funded by the U.S. Department of Energy, Office of Science, Office of Basic Energy Sciences, Materials Sciences and Engineering Division under Contract No. DE-AC02-05-CH11231 within the Quantum Materials Program (KC2202). This work used resources of the Spallation Neutron Source, a DOE office of Science User Facility operated by the Oak Ridge National Laboratory.\\ 
Notice:  This manuscript has been authored by UT-Battelle, LLC, under contract DE-AC05-00OR22725 with the US Department of Energy (DOE). The US government retains and the publisher, by accepting the article for publication, acknowledges that the US government retains a nonexclusive, paid-up, irrevocable, worldwide license to publish or reproduce the published form of this manuscript, or allow others to do so, for US government purposes. DOE will provide public access to these results of federally sponsored research in accordance with the DOE Public Access Plan (http://energy.gov/downloads/doe-public-access-plan).

\item[Contributions]  BF, ML, LC, RS and HHK wrote the paper.   BF, ZY and RF performed the neutron experiments.  BF, ZY, RF, YH and IS analyzed neutron data. RS, SK, FB, VG and HHK conducted NMR and M\"ossbauer experiments. BFran advised BK, AA and RB on performing and interpreting PDF refinement.  ML and SA conceived and performed the theory work.  MF grew the samples. BF conceived of the overall project.  All authors edited and assessed the manuscript.

 \item[Competing Interests] The authors declare that they have no
competing interests.

 \item[Correspondence] Correspondence and requests for materials
should be addressed to B.F.
~(bkfreelon@uh.edu) and M.L. ~(mslaad@imsc.res.in).
\end{addendum}


\end{document}